\documentclass[sigconf]{acmart}

\usepackage{hyperref}

\usepackage{balance}

\def\BibTeX{{\rm B\kern-.05em{\sc i\kern-.025em b}\kern-.08emT\kern-.1667em\lower.7ex\hbox{E}\kern-.125emX}}

\usepackage[noline]{algorithm2e}
\SetKwInOut{Input}{input}
\SetKwInOut{Output}{output}
\newlength\inoutlen

\usepackage{graphicx}
\usepackage{subcaption}
\usepackage{epsf}
\newcommand{\epsin}[2]%
{\setlength{\epsfxsize}{#2\hsize}\centerline{\epsfbox{#1}}}
\renewcommand\footnotetextcopyrightpermission[1]{}

\usepackage{paralist}

\usepackage{flushend}

\begin{document}
\fancyhead{}
\let\url\nolinkurl

\title{Silent Data Corruptions at Scale}

\author{Harish Dattatraya Dixit}
\affiliation{%
\institution{Facebook, Inc.}}
\email{hdd@fb.com}

\author{Sneha Pendharkar}
\affiliation{%
\institution{Facebook, Inc.}}
\email{spendharkar@fb.com}

\author{Matt Beadon}
\affiliation{%
\institution{Facebook, Inc.}}
\email{mbeadon@fb.com}

\author{Chris Mason}
\affiliation{%
\institution{Facebook, Inc.}}
\email{clm@fb.com}

\author{Tejasvi Chakravarthy}
\affiliation{%
\institution{Facebook, Inc.}}
\email{teju@fb.com}

\author{Bharath Muthiah}
\affiliation{%
\institution{Facebook, Inc.}}
\email{bharathm@fb.com}

\author{Sriram Sankar}
\affiliation{%
\institution{Facebook Inc.}}
\email{sriramsankar@fb.com}

\renewcommand{\shortauthors}{Dixit et al.}

\begin{abstract}
Silent Data Corruption (SDC) can have negative impact on large-scale infrastructure services. SDCs are not captured by error reporting mechanisms within a Central Processing Unit (CPU) and hence are not traceable at the hardware level. However, the data corruptions propagate across the stack and manifest as application-level problems. These types of errors can result in data loss and can require months of debug engineering time.

In this paper, we describe common defect types observed in silicon manufacturing that leads to SDCs. We discuss a real-world example of silent data corruption within a datacenter application. We provide the debug flow followed to root-cause and triage faulty instructions within a CPU using a case study, as an illustration on how to debug this class of errors. We provide a high-level overview of the mitigations to reduce the risk of silent data corruptions within a large production fleet.

In our large-scale infrastructure, we have run a vast library of silent error test scenarios across hundreds of thousands of machines in our fleet. This has resulted in hundreds of CPUs detected for these errors, showing that SDCs are a systemic issue across generations. We have monitored SDCs
for a period longer than 18 months. Based on this experience, we determine that reducing silent data corruptions requires not only hardware resiliency and production detection mechanisms, but also robust fault-tolerant software architectures.

\end{abstract}


\maketitle
\renewcommand{\shortauthors}{Dixit \textit{et al.}}

\section{Introduction}
\label{s:introduction}

Facebook infrastructure serves numerous applications like Facebook, Whatsapp, Instagram and Messenger. This infrastructure consists of hundreds of thousands of servers distributed across global datacenters. Each server is made up of many fundamental components like Motherboard, Central Processing Units (CPU), Dual In-line Memory Modules (DIMMs), Graphics Processing Units (GPU), Network Interface Cards (NICs), Hard Disk Drives (HDDs), Flash Drives and interconnect modules. The key unit that brings all these components together is the CPU. It manages the devices, schedules transactions to each of them efficiently and performs billions of computations every second. These computations power applications for image processing, video processing, database queries, machine learning inferences, ranking and recommendation systems. However, it is our observation that computations are not always \textit{accurate}. In some cases, the CPU can perform computations incorrectly. For example, when you perform 2x3, the CPU may give a result of 5 instead of 6 silently under certain microarchitectural conditions, without an indication of the miscomputation in system event or error logs. As a result, a service utilizing the CPU is potentially unaware of the computational accuracy and keeps consuming the incorrect values in the application. This paper predominantly focuses on scenarios where datacenter CPUs exhibit such silent data corruption. We dive deep into a real-world application-level impact of a corruption, the processes used in debugging such corruption, and conclude with detection and mitigation strategies for silent data corruptions. While we present one case study, we have observed several scenarios, data paths and architectural blocks where SDCs manifest, and hence it is a systemic problem that the industry should tackle collectively.

Prior work ~\citep{Baumann}, ~\citep{ARM_R5_VA}, ~\citep{SEP_uC}, ~\citep{IBM}, ~\citep{Bitflipinjection}, ~ \citep{elliott2013quantifying} within this domain focused on soft errors due to radiation or synthetic fault injection. In contrast, we observe that silent data corruptions are not limited to soft errors due to radiation and environmental effects with probabilistic models. Silent data corruptions can occur due to device characteristics and are repeatable at scale. We observe that these failures are reproducible and not transient. Techniques like Error Correction Code (ECC) are beneficial for reducing the error rates in SRAM. However not all the blocks within a datacenter CPU have similar datapath protection. Moreover, CPU SDCs are evaluated to be a one in a million occurrence within fault injection studies. We observe that CPU SDCs are orders of magnitude higher than soft-error based FIT simulations. CPU SDCs occur at a higher rate due to minimal error correction within functional blocks. With increased silicon density and technology scaling ~\citep{trends}, ~\citep{CMOS_trends}, we believe that academic researchers and industry should invest in methods to counter these issues.

Facebook infrastructure initiated investigations into silent data corruptions in 2018. In the past 3 years, we have completed analysis of multiple detection strategies and the performance cost associated. For brevity, this paper does not include details on the performance vs cost tradeoff evaluation. A follow up study would dive deep into the details. In this paper, we provide a case study with an application example of the corruption and are not using any fault injection mechanisms. This corruption represents one of the hundreds of CPUs we have identified with real silent data corruption through our detection techniques.

The rest of the paper is structured as follows: Section ~\ref{s:related_work} provides an overview of related work within this domain. Section ~\ref{s:defect_categories} walks through the different defect categories in silicon design and manufacturing. Section ~\ref{s:application_level_impact_of_silent_corruptions} details a real-world application example of silent data corruption and propagation of corruptions across the stack. Section ~\ref{s:debugging_silent_data_corruptions_at_scale} lists the best practices for root-causing silent data corruptions at scale, and walks through the debugging for the application in the case study. Section ~\ref{s:revisiting_app_failures}, concludes the debug findings and revisits application failure with a deeper understanding of the CPU defect. Section ~\ref{s:detection_mechanisms} provides a high level overview of fleet detection mechanisms that can be implemented to mitigate the risk of silent errors. Section ~\ref{s:software_fault_tolerant_mechanisms} provides a high level overview of software fault tolerant mechanisms for bitflips and data corruptions.

\section{Related Work}
\label{s:related_work}

Previous work within the silent error domain studies the impact of soft errors due to radiation ~\citep{Baumann}, and how environmental factors can lead to soft errors within the system. The study provides error rate observations for a non ECC protected SRAM. This is calculated using a Soft Error Rate (SER) from radiation resulting in an estimated 50000 FIT (Failure-In-Time: One FIT is equivalent to one failure in 1 billion device hours). Hence they recommend using ECC which reduces the error rate by 1000x for SRAMs.

Experiments with bit-flip injection mechanisms in floating point units~ \citep{elliott2013quantifying} have shown the theoretical impact of bitflips within processors. Bitflip injection mechanisms have also been used to compare the performance of processors under benchmarks with synthetic injection and radiation induced bitflips ~\citep{Bitflipinjection}. A 2012 study on silent data corruption in a HPC cluster with 96 nodes ~\citep{IBM_SDC_at_scale} evaluated the impact of soft errors using fault injector and correcting the corruptions with focus on Message Passing Interface (MPI) Protocols. Within the fault injection study, the fault injector ran with a corruption frequency of 1 in 5 million messages to ensure a relatively high likelihood for an injection. Faster corruption frequency of 1 in 2.5 million messages was also included to evaluate the impact of higher occurrence rates on MPI workloads.

Another set of studies evaluate the risk and mitigation strategies of soft error induced faults within microprocessors. A study from ARM ~\citep{ARM_R5_VA} evaluates the vulnerability assessment of soft-errors on ARM Cortex R5 CPUs by breaking down the percentage of sequential logic vulnerable to soft errors which propagate to output ports. In a collaboration study between Intel and University of Michigan ~\citep{SEP_uC} radiation induced soft errors are identified to not reflect a permanent failure. The study captures the essential metrics required for quantifying soft errors, evaluating Failure-In-Time (FIT) and techniques to reduce the soft error rate using process technology, circuit, and architectural solutions. A similar study from IBM targets 114 SDC FIT for Power4 systems ~\citep{IBM}. All these studies evaluate errors as transient or soft indicating the radiation dependent nature of the error.

ECC reduces the error rate for SRAMs but all the datapaths within datacenter CPUs are not protected by ECC. In addition, the FIT models for CPU also derive from soft error probabilities to evaluate robustness, vulnerability assessments and fault tolerance in the above studies. Since datacenter SDCs are observed to be at higher orders of magnitude, it is valuable for us to explore best practices to debug, detect and mitigate SDCs at scale.

\section{Defect Categories}
\label{s:defect_categories}

Each datacenter CPU contains billions of transistors which are switching constantly. These transistors are devices made of chemical compositions predominantly of silicon with p-type and n-type impurities. A CPU is designed to meet the desired computing requirements while keeping within the power, thermal and spatial constraints for the chip. Once the design is signed off, a layout for the chip is prepared where billions of logic gates are placed to minimize electrical noise, crosstalk, boost signal distribution and stability. Finally, after validation of all the functional, architectural, and physical requirements, the chip is taped-out as part of the chip development process. After the manufacturing process, the designed chips are then subject to test patterns for expected functional behavior, quality control and eventually shipped to all the computing customers worldwide.

\subsection{Device Errors}
\label{ss:device_errors}
Within the manufacturing and design process there are opportunities for defects to manifest. It is possible that the design has corner case scenarios. For example, a block which manages the cache controller under a particular power state can have functional limitations. This can result in the device being stuck or manifest functional errors. During placement and routing of blocks within the CPU, there could be uncertainty in the arrival time for signals, which can then lead to an erroneous bit-flip. One example of such failure is a timing path error. While manufacturing, it is also probable that all the transistors are not etched reliably, and all of them do not have the same peak-operating voltage or power thresholds. This can lead to variations in device characteristics and results in manufacturing errors \citep{manufacturing_defects}, \citep{intermittent_faults}.

\subsection{Early Life Failures}
\label{ss:early_life_failures}
Some of the early life failures are identified during manufacturing tests, these failures negatively impact the yield of the process. A few of the devices are healthy enough to pass the manufacturing test pattern but exhibit failure symptoms only after they have been in the field serving workloads. Depending on the type of electrical weakness within the transistor, a fault may manifest within the first weeks, months or any time before the end of the expected device life \citep{ELF}, \citep{ELF2}. These failures are classified as early life failures.

\subsection{Degradation}
\label{ss:degradation}
It is also possible for the devices to get weaker with usage. A computational block used frequently can show wear and tear, and degrade faster than the other parts of the CPU. These are uncommon in comparison to early life failures but are still observed within the industry. An example of this can be seen in another device used in servers - \textit{Rowhammer attacks} for DDR4 memory components \citep{frigo2020trrespass}. Devices incorporate error correction mechanisms like Error Correction Codes (ECC) to protect against degradation within the device. Degradation based failures can have negative impact as the aging is not uniform across different chips that fall under this failure category.

\subsection{End-of-Life Wear-out}
\label{ss:end_of_life_wearout}
When the device has been in the field serving workloads for a while, beyond their rated life, the entire silicon starts exhibiting wear-out \citep{EOL}, \citep{EOL2}, \citep{agarwal2007circuit}. This is observed in most components and is classified as silicon wear-out within the bathtub curve modeling of failures. This is also typically the duration for which the failure analysis support or firmware support exists for CPUs.

\vspace{10pt}

All the four failure modes described above have the potential to lead to SDC within a fleet of machines. It is statistically more likely to encounter silent data corruption with increasing CPU population. It is our observation that increased density and wider datapaths increase the probability of silent errors. This is not limited to CPUs and is applicable to special function accelerators and other devices with wide datapaths. In the next section, we analyze how these errors propagate across the stack and cause application-level manifestations. We present ways to debug them at scale and discuss detection practices at different abstraction levels.

\section{Application Level Impact of Silent Corruptions}
\label{s:application_level_impact_of_silent_corruptions}

Facebook infrastructure is made up of hundreds of thousands of servers and has billions of users accessing our applications. With billions of users accessing the Facebook family of applications, the infrastructure receives billions of requests per day. With billions of user queries, image uploads, and media content, the processing required for these applications needs to be fast, reliable, and secure. We utilize fundamental concepts within distributed systems to partition our applications and optimize each of the said partitions. A typical application can require anywhere between tens of machines to hundreds of thousands of machines based on the complexity, resource profile and computing needs of the application. One such partition is our querying infrastructure. This querying infrastructure is used to fetch and execute SQL and SQL like queries (Presto, Hive, Spark) \cite{MySQL}, \cite{Spark} across multiple datasets.

\begin{figure}[tbhp]
\includegraphics[width= 8cm]{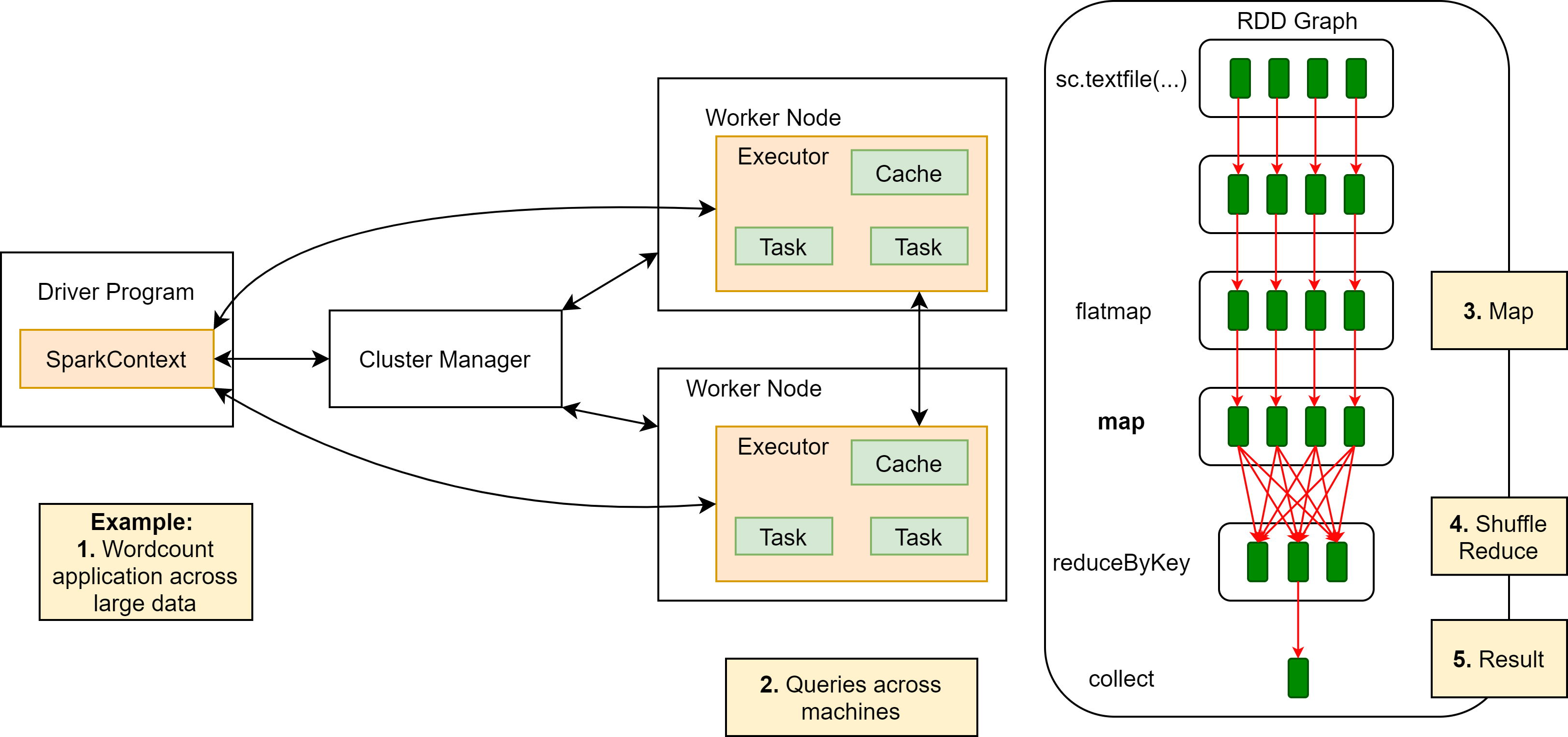}
\caption{High Level Spark Architecture}
\label{f:spark_architecture}
\end{figure}

\subsection{Spark}
\label{ss:spark}

Figure \ref{f:spark_architecture} \cite{Cluster_Overview} describes a typical architecture of a spark cluster. Spark is a widely known distributed processing framework which works based on the concept of Resilient Distributed Datasets (RDDs) each of which can be run in parallel. The results for a large data processing application are produced after several key steps. At a high level, a mapping function first maps the data blocks. This is followed by a reduction operation which aggregates the results across multiple RDDs. The result is presented in the collect phase after reduction.

For example, a Wordcount application, trying to count the number of occurrences of each word within a large file would execute in the following way. The large file would be split into multiple RDDs. The RDDs are assigned to worker nodes, these worker nodes compute the word-count for a subset of the dataset. Results from each node are aggregated together in the shuffle reduce stage. Finally, an output table of each word and its associated occurrence count is provided to the user. In a large infrastructure environment like Facebook, these applications run millions of such computations every day.

\subsection{FB Compression Application}
\label{ss:fb_compression_application}

\begin{figure}[tbhp]
\includegraphics[width= 8cm]{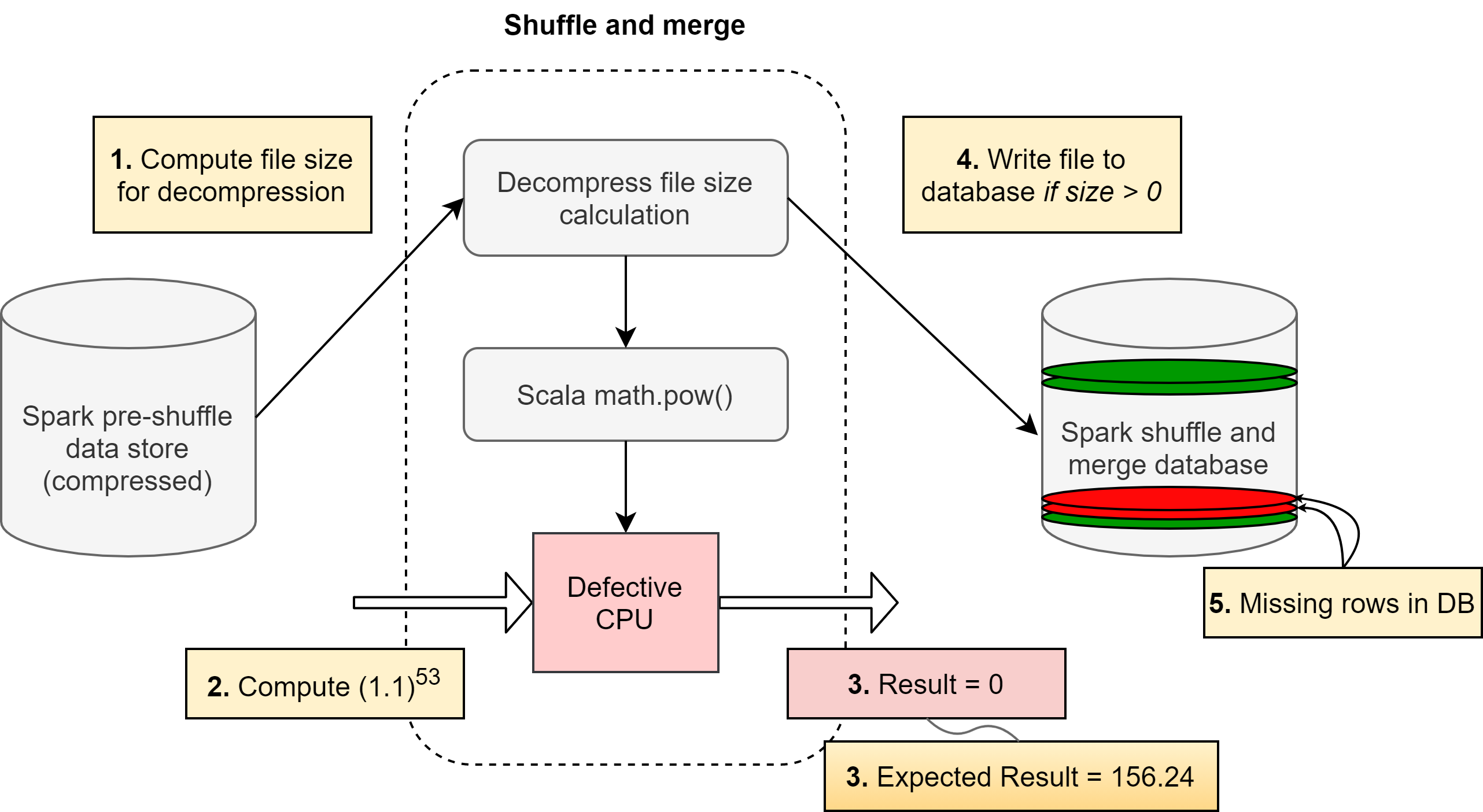}
\caption{Application level silent data corruption}
\label{f:app_sdc}
\end{figure}

Like wordcount, compression is a technique which is used to reduce the storage footprint of datastores and can make use of the spark architecture. There are multiple algorithms for compression. In this paper we will not be going into details of the algorithms. Interested readers can review the following papers for details and comparison of compression algorithms \citep{sayood2017introduction}, \citep{bhattacharjee2013comparison}, \citep{kodituwakku2010comparison}. Files are usually compressed when they are not being read and decompressed when a request is made for reading the file. In a large infrastructure, millions of compression and decompression operations are performed every day. In this example, we are mainly focusing on the decompression aspect of files. We have a database, where the files are compressed and stored within a data store. Upon request, multiple sets of these files are sent to the decompression pipeline. Before a decompression is performed, file size is checked to see if the file size is greater than 0. A valid compressed file with contents would have a non-zero size. Figure \ref{f:app_sdc} shows the manifestation of corruptions and interlink to the database pictorially.

In one such computation, when the file size was being computed, a file with a valid file size was provided as input to the decompression algorithm, within the decompression pipeline. The algorithm invoked the power function provided by the Scala library (Scala: A programming language used for Spark) \cite{Scala}. Interestingly, the Scala function returned a \textbf{0} size value for a file which was known to have a non-zero decompressed file size. Since the result of the file size computation is now 0, the file was not written into the decompressed output database.

Imagine the same computation being performed millions of times per day. This meant for some random scenarios, when the file size was non-zero, the decompression activity was never performed. As a result, the database had missing files. The missing files subsequently propagate to the application. An application keeping a list of key value store mappings for compressed files immediately observes that files that were compressed are no longer recoverable. This chain of dependencies causes the application to fail. Eventually the querying infrastructure reports critical data loss after decompression. The problem's complexity is magnified as this manifested occasionally when the user scheduled the same workload on a cluster of machines. This meant the patterns to reproduce and debug were non-deterministic.

\section{Debugging Silent Data Corruptions at Scale}
\label{s:debugging_silent_data_corruptions_at_scale}

With concerted debugging efforts and triage by multiple engineering teams, logging was enabled across all the individual worker machines at every step. This helped narrow down the host responsible for this issue. The host had clean system event logs and clean kernel logs. From a system health monitoring perspective, the machine showed no symptoms of failure. The machine sporadically produced corrupt results which returned zero when the expected results were non-zero.

The reproducer at a multi-machine querying infrastructure level was then reduced to a single machine workload. From the single machine workload, we identified that the failures were truly sporadic in nature. The workload was identified to be multi-threaded, and upon single threading the workload, the failure was no longer sporadic but consistent for a \textit{certain subset of data values} on one particular core of the machine. The sporadic nature associated with multi-threading was eliminated but the sporadic nature associated with the data values persisted. After a few iterations, it became obvious that the computation of

\[ Int(1.1^{53}) = 0 \]
as an input to the \textit{math.pow} function in Scala would always produce a result of 0 on Core 59 of the CPU. However, if the computation was attempted with a different input value set
\[ Int(1.1^{52}) = 142 \]
the result was accurate.

The next step in the process was to gain a deeper understanding of the scenarios the corruptions manifest in. Any other variants associated with this silent data corruption also require investigation. To confirm the data dependency of the issue, we ran multiple iterations on Core 59. Following shows an example of 3 iterations where 2 of the computations produce faulty results repeatedly.


\hfill\\
\noindent\textbf{Core pinned Scala workload}

\begingroup
\fontsize{7pt}{9pt}\selectfont
\begin{verbatim}

[root@hostname ~]#
for x in {0..2}; do taskset -c 59 ./bitflip_repro.sh; done
# Int(1.1^{53}), Int(1.1^{68}), Int(1.1^{78})

Iteration 1: 0, 0, 1692
Iteration 2: 0, 0, 1692
Iteration 3: 0, 0, 1692

\end{verbatim}
\endgroup

The data dependency is clearly established for the defect. In this example, core 59 is faulty. Ideally when workloads are faulty, the workload can be stepped through GNU Project debugger (GDB) \cite{GDB} and reverse engineered. The instruction data could be compared to a reference computation by stepping through instructions. This step-through process, while time-consuming, enables debugging of silent errors. However, Scala is a language whose workloads cannot be stepped through in GDB. Scala is compatible to run Java Byte Code in a Java Virtual Machine (JVM). Java Byte Code (JBC) \cite{Java_Byte_Code} is compiled by a Just-In-Time (JIT) compiler.

\subsection{Tools}
\label{ss:tools}

We need to perform language conversion while keeping reproducer consistency to triage the root-cause. In this example, we traverse from Scala language reproducer to Java reproducer to JIT compiled JBC to Assembly to triage the instruction level root-cause and enable the reproducer code. Unlike C and C++, Just-In-Time (JIT) compiled code is not compiled ahead of time. However, to debug a silent error, we cannot proceed forward without understanding which machine level instructions are executed. We either need an ahead-of-time compiler for Java and Scala or we need a probe, which upon execution of the JIT code, provides the list of instructions executed.

\subsubsection{Example Scala to Java Byte Code}
\label{ss:scala_to_jbc}
\hfill\\

\noindent The first step to get to assembly is to convert the reproducer from Scala to Java. There are more resources to aid this conversion. We can use the Scala compiler (scalac) to obtain the Java Class routines for the source code. To obtain the Scala compiled Java Byte Code, we modified the Scala script to a Scala compiler friendly reproducer code.

\begingroup
\fontsize{7pt}{9pt}\selectfont
\begin{verbatim}
[root@hostname ~]# scalac Bitflip.scala
# This generates the intercompatible scala/java class files
# This can be read as Java Byte Code.
[root@hostname ~]# javap -c -v Bitflip\$.class
\end{verbatim}
\endgroup

\subsubsection{GCJ}
\label{ss:gcj}
\hfill\\

\noindent GCJ \cite{GCJ}{} was an open source ahead-of-time compiler which could convert JBC to blobs of object files and binary. This binary can be used within GDB to debug. However, the tool development has been deprecated since 2008, and CentOS deprecated the tool in 2010. Without an \textit{ahead-of-time} compiler, it is challenging to perform the static conversion of Java Byte Code to assembly.

\subsubsection{HotSpot}
\label{ss:hotspot}
\hfill\\

\noindent Java provides options to use \textit{+PrintAssembly} to act as a probe and print assembly of the executed code with the use of HotSpot Profiling. To support \textit{+PrintAssembly}, there are 2 requirements,
\begin{itemize}
\item \textbf{Virtual machine with support for hotspot profiler:} This can be identified for an example machine using the following command. An output providing HotSpot confirms that the virtual machine enables profiling. Version numbers shown here are example versions and are not representative of any deployment.

\begingroup
\fontsize{7pt}{9pt}\selectfont
\begin{verbatim}
$> java -version
java version "A.B.C_DEF"
Java(TM) SE Runtime Environment (build G.H.I_JKL-MNO)
Java HotSpot(TM) 64-Bit Server VM (build PQ.RST-UVW, mixed mode)
# This means the VM can be profiled.
\end{verbatim}
\endgroup

\item \textbf{Library for profiling:}
Hotspot is a performance profiler used to analyze hot spots for a program. These hotspots are optimized for high performance execution with minimal overhead for the less-performance critical code. The profiler enables the option for \textit{PrintAssembly} \cite{HotSpot}, and can print the assembly compiled by JIT. These assembly instructions subsequently enable us to root cause and triage the failing instruction.

\end{itemize}
After enabling the profiler, we obtain the assembly that the code executes (JIT + Hotspot output assembly). Our first version of the assembly was 430K lines. With our assembly, we can debug the silent error. The Scala \textit{math.pow} functions are identified within the 430K line assembly. We parse the 430K line assembly to optimize the reproducer. However, the disassembly does not output the sequence of executed instructions but rather lists the methods used in the call stack. The sequencing can be unclear. To obtain a reproducer, we need to sanitize, reverse engineer with a smaller assembly code. From this raw assembly, we can understand the sequence of instructions sent to the CPU and root-cause the faulty instruction by following the best practices to debug silent errors. 

\subsection{Best Practices for Silent Error Debug}
\label{ss:best_practices_for_silent_error_debug}

A few guidelines while reverse engineering the printed assembly code. While these guidelines are derived from this example, they can be leveraged for debugging similar silent data corruptions. 

\begin{itemize}
\item \textbf{Absolute address references:} Leaving absolute addresses to jump to within the code while optimizing for a reproducer will lead to segmentation faults. Instead of managing all the memory locations, it is preferred to eliminate the absolute address reference if that section of assembly is found to have no dependency on the reproducibility.

\item \textbf{Unintended branches:} If unintended branch and jump calls are left unmapped, the code crashes with segmentation faults and undefined code branches. This introduces more variability within the function. It is advisable to limit variability when attempting for a deterministic bitflip reproducer.

\item \textbf{External Library References:} Identify which instructions invoke a call outside the current code path to external libraries. With the goal of a minimal reproducer, it is preferred to not have external library dependencies.

\item \textbf{Compiler Optimization:} High performance code features multi-pass compiler optimizations. Observing optimization to mathematical equations can help in understanding the critical assembly required for the reproducer. Optimizations may not be intuitive while stepping through assembly instructions.

\item \textbf{Stub and Redundant Instructions:} It is preferred to eliminate redundant and stub instructions. Stubs are used by Scala for book-keeping and are not relevant for debugging the failing instruction. Stub instructions do not interfere with functionality outside of the Scala execution context.

\item \textbf{Input/Output registers:} For any bitflip reproducer we need to identify the data input and result registers for the critical instructions. After identification, additional instructions must be added to provide user inputs and obtain results. This enables a stable reproducer code and enables identification of data dependency for the silent data corruption.

\item \textbf{Managing Stack Frames:} Standalone assemble reproducers require stack frames to be appropriately managed. Managing transactions into the stack frame to prevent buffer overflow or underflow is critical for stability. Without stack frames, reproducer code cannot manage stack-based requests or function calls.

\item \textbf{Memory-offset references:} Registers typically use memory offsets within instructions. The offsets must be initialized appropriately. If offsets are not calculated and initialized, we will encounter segmentation faults or reproducer corruption due to uninitialized data.

\item \textbf{Special Function Units:} We need to monitor transactions to special function units (like ALU, DSP, FPU, AVX etc) as they bring in approximations. In addition, special function units utilize varied bit widths, special function registers and stack architecture.

\item \textbf{Main Frames:} A standalone reproducer cannot be complete without appropriate main frames and function frames. This makes the code executable.
\end{itemize}

In this section, we are purely focusing on the best practices for silent error debugging, and not on the knowledge prerequisites about CPU architectures or GDB internals.
\begin{itemize}
\item We are skipping over the hardware architecture and implementation details for all the CPU sub-blocks. Details associated with the status flags, differences between special function stacks and normal integer stack, instruction truncation and handshakes between different precision bit-width and operand types are skipped. All of these are key to identify the steps within a CPU and are widely documented in published research.
\item We are skipping over all the steps within GDB, and the methods to print, step through commands, scripting through different stacks, registers, memory addresses as these are documented widely.
\end{itemize}

After reverse engineering, identifying the handshake between hardware blocks and dependency graphs for assembly, we can arrive at a simpler reproducer. Here are some interesting observations from the assembly that were obtained for this example.

\begin{itemize}
\item For squaring a number, the scala compiler implements a fast optimization using look-up tables.
\item \textit{math.pow} function is in-lined in the power function, even though PrintAssembly prints them separately.
\item Scala \textit{math.pow} computes powers using the formula -
\begingroup
\fontsize{10pt}{9pt}\selectfont
\[ x^y = 2^{y * log_2x} \]
\endgroup
\end{itemize}

We step-through instructions in GDB. During the step-through process, instruction operands, memory and register states, and instruction outputs are examined for corruption. As a result of this process, we obtain the faulty instruction within the defective CPU. 

\subsection{Assembly Level Test Case}
\label{ss:assembly_level_test_case}


\begin{figure}[thbp]
\includegraphics[width= 9cm]{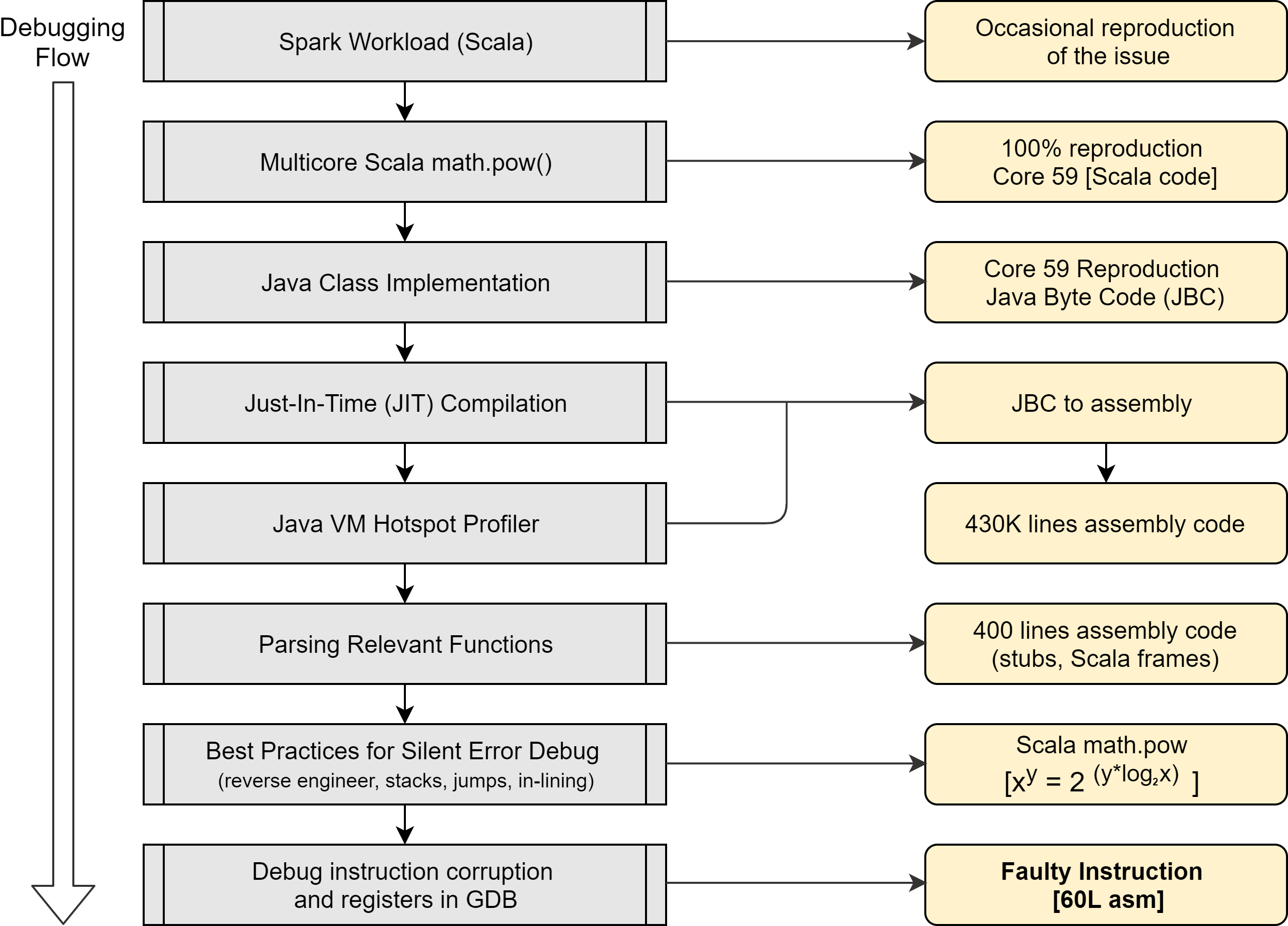}
\caption{High Level Debug Flow}
\label{f:high_level_debug_flow}
\end{figure}

Once the reproducer is obtained in assembly language, we optimize the assembly for efficiency. The assembly code accurately reproducing the defect is reduced to a 60-line assembly level reproducer. We started with a 430K line reproducer and narrowed it down to 60 lines. Figure ~\ref{f:high_level_debug_flow} provides a high level debug flow followed for root-causing silent errors.

\section{Revisiting Application Failures}
\label{s:revisiting_app_failures}
Note that that all the machines operating the application do not have any logs or system level health information indicating this failure mode. We identified cases of corruption impacting computations involving non-zero operands and results. For example, the following incorrect computations were performed on the defective CPU. We identified that the computation affected positive and negative powers for specific data values. In some cases, the result was non-zero when it should have been zero. We noticed incorrect values with varying degrees of precision.

\vspace{10pt}

\noindent\textbf{Example errors:}
\begingroup
\fontsize{7pt}{9pt}\selectfont

\[ Int[ (1.1)^3 ] = 0, expected = 1. \]
\[ Int[ (1.1)^{107} ] = 32809, expected = 26854. \]
\[ Int[ (1.1)^{-3} ] = 1, expected = 0. \]

\endgroup

As a result, an application could have decompressed files of incorrect size and are incorrectly truncated without an End-Of-File (EoF) terminator. This leads to dangling file nodes, missing data, and no traceability of a corruption within an application. The intrinsic data dependencies on the core as well as the data inputs make the corruptions close to impossible to detect and root-cause without a targeted reproducer. This is challenging, especially in a scenario where a fleet has hundreds of thousands of machines performing a few million computations every second. We identified additional machines with the targeted reproducer. We integrated our lessons from the reproducer into detection mechanisms within the fleet. In addition, the best practices identified for silent error debugging enable faster root-cause and sensitivity analysis for similar errors within the fleet.

We initiated efforts in estimating the business impact due to SDCs by quantifying the scale and criticality of the problem to our infrastructure. Given the silent nature of these errors, evaluating the scale of the problem was challenging at first. Initially the calculations for defective-parts-per-million predictions, debug engineering time allocations and business impact were based on heuristics and smaller datasets. With data collection and analysis in the past 18 months, we arrived at empirical values and ranges for each of the above.

\subsection{Hardware approaches to counter SDCs}
\label{ss:hardware_approaches}

We observe that silent data corruptions are not limited to rare one in a million occurrences within a large-scale infrastructure. These errors are systemic and are not as well understood as the other failure modes like Machine Check Exceptions. There are several studies evaluating the techniques to reduce soft error rate within processors ~\cite{weaver}, ~\cite{shirvani}, we can extend these techniques to repeatable SDCs which can occur at a higher rate. We can mitigate the exposure of applications to silent errors by using different strategies.
\begin{itemize}
\item \textbf{Protected Datapaths:} Augmenting blocks within the device to have increased datapath protection using algorithms similar to Error Correcting Codes (ECC) can increase resiliency of the device.
\item \textbf{Specialized Screening:} Dedicated screens and test patterns within the manufacturing flow to detect silent errors. Testing with randomized data streams can increase the probability of hit rate within manufacturing testing.
\item \textbf{Understanding @Scale Behavior:} Close partnership with the customers using devices at scale to understand and evaluate the impact of silent errors. It is beneficial to study occurrence rates, time to failure in production, dependency on frequency, voltage, and environmental conditions to obtain insights into manifestations of SDCs.
\item \textbf{Architectural priority:} With increased density, wider datapaths and technology scaling; we are more likely to observe silent data corruptions moving forward. Prioritizing protection against silent data corruption within our architectural choices can enable future semiconductor devices to be more resilient. 

\end{itemize}
The strategies described above are not limited to CPUs and can be extended to Application Specific Integrated Circuits (ASIC) and devices with wider data paths and unprotected logic.

\section{Detection Mechanisms}
\label{s:detection_mechanisms}

To detect errors of this type in the fleet, we need workloads which execute specific types of computations. We then compare the results of these computations with known reference values to ensure that the results are accurate. Silent corruptions tend to be data dependent making it difficult to predict their occurrence in the fleet. Given that any downtime for testing in a production fleet is an efficiency loss, this can be achieved in 3 different ways:

\subsection{Opportunistic}
\label{ss:opportunities}
Opportunistically utilize machines in maintenance states and perform instruction level accuracy validation with randomized data inputs. The challenge here is that the coverage of the fleet is based on how frequently machines fall into these opportunistic states. In a large fleet, we do not expect large percentages of machines to be in these states, however there are transition states (provisioning, service setup etc) that can be used opportunistically.

\subsection{Periodic}
\label{ss:periodic}
Implement a scheduler which periodically monitors machines for silent error coverage and then schedules machines based on a periodic timer (for example: 15 days) for testing. Here the overhead is high as the machine is forced to an out of production status to perform testing at a specified schedule.
\subsection{Production Friendly}
\label{ss:production_friendly}
Tests can be optimized to be minimal in size and run-time. This can enable test instructions to be executed in parallel with the workloads on the machine. The result is sent to a collector to notify a pass or fail status for the machine. This method requires close coordination with the workload to not have any adverse impact on the production workload.

\section{Software Fault Tolerant Mechanisms}
\label{s:software_fault_tolerant_mechanisms}

To deal with silent errors, we need to rethink the robustness of infrastructure software design philosophies and software abstractions.

\subsection{Redundancy}
\label{ss:redundancy}

A better way to prevent application-level failures is to implement software level redundancy and periodically verify that the data being computed is accurate at multiple checkpoints. This is a tried and tested method implemented in space research ~\citep{sklaroff1976redundancy}, aircraft ~\citep{FRANK1990459} and automobiles ~\citep{Automobile_redundancy}. It is important to consider the cost of accurate computation while adopting these approaches to large-scale data center infrastructure. The cost of redundancy has a direct effect on resources, more redundant the architecture, the larger the duplicate resource pool requirements. However, this provides probabilistic fault tolerance to the application.
\subsection{Fault Tolerant Libraries}
\label{ss:fault_tolerant_libraries}

Adding fault tolerance into well-known open-source libraries like PyTorch would greatly aid the applications to prevent exposure to silent data corruptions. Building algorithmic fault tolerance adds additional overhead on the application. This can be implemented with negligible drop in performance. This effort would need a close handshake between the hardware silent error research community and the software library community.

Facebook infrastructure has implemented multiple variants of the above hardware detection and software fault tolerant techniques in the past 18 months. Quantification of benefits and costs for each of the methods described above has helped the infrastructure to be reliable for the Facebook family of apps. A subsequent publication will go into statistical detail on trade-offs across detection strategies and coverage scenarios for detection mechanisms and fault tolerant software libraries.

\section{Conclusions}
\label{s:conclusion}

Silent data corruptions are \textbf{real} phenomena in datacenter applications running at scale. We present an example here which illustrates one of the many scenarios that we encounter with these data dependent, reclusive and hard to debug errors. Understanding these corruptions helps us gain insights into the silicon device characteristics; through intricate instruction flows and their interactions with compilers and software architectures. Multiple strategies of detection and mitigation exist, with each contributing additional cost and complexity into a large-scale datacenter infrastructure. A better understanding of these corruptions has helped us evolve our software architecture to be more fault tolerant and resilient. Together these strategies allow us to mitigate the costs of data corruption at Facebook's scale.

\vspace{20pt}
\textbf{Acknowledgement}
The authors would like to thank Manish Modi, Vijay Rao, T.S. Khurana, Aslan Bakirov, Melita Mihaljevic, Kushal Thakkar, Nishant Yadav, Aravind Anbudurai, Jason Liang, Jianyu Huang, Sihuan Li, Jongsoo Park and other infrastructure engineers for their inputs in the implementation of solutions and valuable technical suggestions.

\bibliographystyle{ACM-Reference-Format}
\bibliography{silent_error.bib}

\end{document}